\documentclass[pra,reprint,onecolumn,floatfix,notitlepage,showkeys]{revtex4-1}

\usepackage{graphicx}
\usepackage{amssymb}
\usepackage{epsfig}
\usepackage{epsf}
\usepackage{epstopdf}
\usepackage{xcolor}

\begin{document}

\title{Fuzzy Q-Learning Based Multi-Agent System for Intelligent Traffic Control by a Game Theory Approach}

\author{Abolghasem Daeichian}
\email{ E-mail:a-daeichian@araku.ac.ir, a.daeichian@gmail.com}
\thanks{This is a pre-print of an article published in Arabian Journal for Science and Engineering. The final authenticated version is available online at: https://doi.org/10.1007/s13369-017-3018-9}
\affiliation{Department of Electrical Engineering, Faculty of Engineering, Arak University, Arak, 38156-8-8349 Iran}
\author{Amir Haghani}
\affiliation{Department of electrical engineering, Payam Institute of Higher Education,Golpayegan, Isfahan, Iran}

\begin{abstract}
This paper introduces a multi-agent approach to adjust traffic lights based on traffic situation in order to reduce average delay time. In the traffic model, lights of each intersection are controlled by an autonomous agent. Since decision of each agent affects neighbor agents, this approach creates a classical non-stationary environment. Thus, each agent not only needs to learn from the past experience but also has to consider decision of neighbors to overcome dynamic changes of the traffic network.
Fuzzy Q-learning and Game theory are employed to make policy based on previous experiences and decision of neighbor agents. Simulation results illustrate the advantage of the proposed method over fixed time, fuzzy, Q-learning and fuzzy Q-learning control methods.
\end{abstract}
\maketitle

\section{Introduction}	
Urbanization, increasing number of vehicles, and lack of transport infrastructures have increased travel time, fuel consumption, and air pollution. Therefore, urban life equals with waste of time, less clean air, and acoustic pollution. Conventional fixed traffic management systems are not able to fight complexity and dynamic of large traffic networks. While, artificial intelligence (AI) are greatly employed to develop intelligent traffic systems (ITS) ~\cite{kponyo2016distributed,BGS2010IET,BK2014KER,R2014AJSE}, 
multi-agent system is an approach to model ITS ~\cite{RPM204PH,vilarinho2017intelligent}. This framework consists of a population of intelligent and autonomous agents work together in an environment ~\cite{schaefer2016multi}. Traffic lights ~\cite{L2007IJCSNS}, vehicles ~\cite{ASM2005TRPB}, and pedestrians ~\cite{T2006TRPF} are considered as agents in modeling of urban traffic networks. 
Each agent needs to learn from the past experiences which is a key point to approximate a better decision-making policy. Multi-agent model-based ~\cite{W2000CP} as well as model-free ~\cite{CBK2011IJSSS} reinforcement learning (RL) techniques are widely used in researches on ITS ~\cite{PB2011ITS,BGS2010IET}. 

In a multitude of researches, any agent only considers its own traffic state in order to determine the control policy. For example, single intersection with two phases is investigated in ~\cite{abdulhai2003reinforcement}. Length of vehicles queue waiting on the light is considered as state which can be measured by the agent. It decides on extend green time or change it to the next phase so that the number of vehicles waiting on the light is minimized. The results show superiority of Q-learning agent over uniform traffic flows and constant-ratio traffic flows.
In ~\cite{W2000CP}, traffic lights are considered as agents which communicate with vehicles. The vehicles estimate their mean waiting time and transmit this time to traffic light where a popular RL algorithm, namely Q-learning, is used to provide a control for traffic signal scheduling. Results of this study show 22\% reduction in waiting time compared to constant time lights. 
Multi-objective reinforcement learning is utilized to control several traffic lights in ~\cite{houli2010multiobjective}. Optimization goals include number of stops of a vehicle, mean stopping time, and length of vehicles' queue on the next intersection. Its results indicate that multi-RL can effectively prevent the queue spillovers under congested condition to avoid large-scale traffic jams.
Bull et.al. used learner classifiers to control light traffic including 4 intersections ~\cite{bull2004towards}. In this research, traffic lights include two phases at each intersection, where one phase is for moving north-south and one is for east-west. Controller at each intersection, obtains optimum phase time through extracting if-then rules. Its results show that performance of the traffic light using learner classifier system has improved significantly compared to constant time traffic light.
In ~\cite{SSP2005C}, the learning purpose is modeled in such a way that states indications are based on the summation of the cars waiting times. Obviously the more cars information is received, the model will be more complicated and state space will be larger. This issue is one of the significant problems of large networks. 
Adaptive control, which is introduced in ~\cite{PB2011ITS}, uses the approximate of a function as mapping of states to scheduling. 
Fuzzy inference engine is exploited to decrease systematic faults of Q-algorithm in ~\cite{PR2010ITSC}. The results demonstrate that not only learning in fuzzy framework is done faster than Q-learning but also delay in intersections is decreased considerably. A multi-agent fuzzy approach is proposed in \cite{iyer2016intelligent}, where Q-learning updates the set of rule base in fussy inference engine.
In ~\cite{DBP2006ACM} a new method which has the capability to estimate an incomplete model of environment is described for a given non-static environment. This method is applied in a network composed of 9 intersections. The reported results show that this method has better performance than the model-free methods and model-based methods, but could not be generalized and used in larger networks.

In other researches, agents consider other agents in determination of their own control policy. For instance, coordination among agents is desired in ~\cite{medina2010arterial} where the agents not only consider number of waiting vehicles on its own intersection but also they consider number of vehicles which have stopped in adjacent intersections. The RL is applied on 5 intersections within three different scenario. The overall results show improvement in delay time.
In ~\cite{W2000CP}, RL is used to control the traffic in a grid where a type of cooperative learning simultaneously controls the traffic signals and determines the optimal routes. One of the main drawbacks of this method is the high costs of communication and information exchange, specifically when intersections of network are increased.
Cooperative RL tries to extract the knowledge from neighbor agents in a scheduling learning ~\cite{SCA2008ICS}. This method is implemented in an area of Dublin including 64 intersections.

This paper introduces a hybrid fuzzy Q-learning and Game theory method for control of traffic lights in multi agent framework. It exploits the benefits of fuzzification as well as interaction with other agents. The traffic network is modeled by considering an autonomous agent controls in which each intersection decides on duration of green phase. The number of vehicles in different inputs of the intersection are measured by the corresponding agent. Any agent interacts with neighbour agents by getting a reward from each decision. 
This paper proposes that each agent fuzzify the inputs and utilizes in a fuzzy inference system for fuzzy estimation of traffic model states. The agent uses a Q-learning approach modified by Game theory to learn from the past experiences and consider the interaction with neighbor agents.
The agent gets a reward proportional to its own traffic state and a reward from each decision from neighbour agents to update its Q-learning algorithm. The neighbour reward and its weighting in Q-value update is proposed to be fuzzy in the proposed method.
The proposed method is applied on a five-intersection traffic network. The simulation results indicate that proposed method outperforms the fixed time, fuzzy, Q-learning and fuzzy Q-learning control methods in the sense of average delay time.

This paper is unfolds as follows. After this introduction, Q-learning and its fuzzy version are described in the next section. Section 3 is devoted to application of Game theory in ITS. Section 4 and 5 are about problem statement and proposed solution, respectively. Simulation results are given in section 6. Finally, the paper is concluded in section 7.

\section{Q-learning and fuzzy Q-learning}
The objective of agents which act in dynamic environments is making optimum decisions. If the agents are not aware of rewards corresponding to various actions, selecting a proper action would be challenging. To achieve this goal, learning adjusts agents' action selection based on collected data. Each agent tries to optimize its actions with dynamic environment via trial and error in reinforcement learning (RL). The RL is actually how different situations are mapped upon actions to receive the best results or the highest reward. In many cases, actions influence the reward of next steps as well as affect the reward of its corresponding step. There are model-based ~\cite{W2000CP} as well as model-free ~\cite{CBK2011IJSSS} RL techniques. In model-free RL, the agent does not need explicit modeling of the environment because its actions could be directly selected based on rewards.
Q-learning is a model-independent approach where the agent does not access to transfer model ~\cite{WD1992ML,AMB2011ICIT}. Suppose that the agent is in a state $s$, performs an action $a$, from which it gets the rewards $r$ from the environment and the environment changes to state $s'$. This is given by a tuple in the form of $(s,a,r,s')$. State-action value which represents the expected total reward resulting from taking action $a$ in state $s$ is denoted by Q-value $Q(s,a)$. The agent starts with random value and after each action they receive a tuple in the form of $(s,a,r,s')$. For each tuple the value of state-action could be calculated according to the following equation:
\begin{eqnarray}\label{QLearning}
    Q(s,a)&=&(1-\alpha)Q(s,a)+\alpha[r+\gamma maxQ(s',a')-Q(s,a)]
\end{eqnarray}
where $\alpha\in[0,1]$ is the learning rate of agent. $\alpha=1$ means that merely new information is considered and zero means that the agent does not have any learning. $\gamma\in[0,1]$ is discount factor which determines future rewards. Zero value for this factor makes the agent opportunist which means that the agent only considers current reward. On the other hand, $\gamma=1$ means that the agent will wait for a longer time to achieve a large reward. Q-learning will converge to optimum value $Q^*(s,a)$  with probability of one if all state-action pairs are experienced repetitively and learning rate decrease during the time ~\cite{PR2010ITSC}.
Generally, RL is useful for solving problems with small dimension discrete state and action space. When the dimension of state and action space becomes larger, the size of search table will be so large that it makes the algorithm very slow due to computational time. On the other hand, when the states or actions are stated continuously, using search table will not be possible. To tackle this problem fuzzy theory is employed. If the intelligent agent has a proper fuzzy set as expert knowledge about the desired area, the ambiguity could be resolved. Thus, intelligent agent can understand vague objectives and unknown environment. In practice, the action in large spaces is facilitated by eliminating Q-values table. In this method everything is based on quality values and fuzzy inference. Fuzzy inference system (FIS) deals with input and Q-learning algorithm uses the follower section and its active rules as states. Reward signal of Q-algorithm is built in accordance with fuzzy logic, environment reward signal and performance estimation of current action. It is tried to select the action which maximizes the reward signal \cite{G2005CS,BLM2009FSS}. Learning system is able to select one action among $j$ actions for each rule. $j$-th possible action in $i$-th rule is denoted by $a[i,j]$ and its value is shown by $q[i,j]$ consider the following rules ~\cite{BLM2009FSS}:
\begin{eqnarray}\label{FuzzyQLearningRule}
  If\  x\  is\  s_i\ &then&\  a[i,1]\  with\  q[i,1] \nonumber\\
  &or&\    a[i,2]\  with\  q[i,2] \nonumber\\
  &\vdots&\\
  &or&\    a[i,j]\   with\  q[i,j]\nonumber
\end{eqnarray}
Learning should find the best result for each rule. If the agent selects an action which results in high value, it may learn optimum policy. Thus, fuzzy inference system may obtain necessary action for each rule ~\cite{BLM2009FSS} .

\section{Game theory in ITS}
Relation between agent oriented environments and games theory originates from the fact that each state of agent-oriented environments can be resembled to a game environment. Profit function of players would be current state of the environment and goal of players is to move toward balanced or equilibrium point (reaching the best decision making policy). 
Some scholars have studied the application of Game theory to control of traffic lights \cite{goyal2017intelligent,groot2017hierarchical}. They integrate Game theory into the multi-agent interaction approach. Some of them suit the traffic problem into a rigorous mathematical game model ~\cite{bell2000game,chen1998game,APM2008CDC} while others modify the learning method of agents based on Game theory ~\cite{XL2009IFCSTA}. 
In ~\cite{APM2008CDC}, signalized intersections are modeled as finite controlled Markov chains and each intersection is seen as non-cooperative game where each player try to minimize its queue. The solutions are given as Nash equilibrium and Stackelberbg equilibrium and the simulation results indicate shorter queue length than adaptive control.
In ~\cite{bell2000game}, a two-player non-cooperative game is articulated between user seeking a path to minimize the expected trip cost and choosing link performance scenarios to maximize the expected trip cost. It shows that the Nash equilibrium point measures network performance.
Intelligent traffic control is expressed as a Cournot game where the traffic authority and the users choose their strategies simultaneously and as a bi-level Stackelberg game where the traffic authority is the leader which determines the signal settings in anticipation of the user reactions.
In ~\cite{XL2009IFCSTA}, Game theory is used to address coordination between agents based on traffic signal control with Q-learning. It specifies strategies ($C(m)=$\{red light time plus 4sec, red light time plus 8sec, red light time minus 4s, red light time minus 8s,unchangeably\}) and actions ($S(n)=$\{east west straight and right turn, south north straight and right turn, east west left turn, south north left turn\}). Then, an interaction mathematical model via Game theory as a four parameter group $G=\{B,A,I,U\}$ is presented. $B$ is a group of decision-makers as players. $A$ is a group of any possible strategies and actions, i.e. $A=C(m)*S(n)$. $I$ represents the information which agents masters. $U$ is the benefit function which adopts Q-value. So, the Nash equilibrium is ~\cite{XL2009IFCSTA}:
\begin{eqnarray}
U_i(a_i^*,a_{-i}^*)&\geq&U_i(a_i,a_{-i}^*)
\end{eqnarray}
where $a_i$ and $a_{-i}$ denote action of $i$-th agent and actions of other agents, respectively. $a_i^*$ and $a_{-i}^*$ represent the actions at Nash equilibrium.
The renewed Q-values in distributed reinforcement Q-learning is used to build the payoff values. Q-value function is updated as:

\begin{eqnarray}\label{GQLearning}
Q_i(s_i,a_i )&=&(1-\alpha_i)Q_i(s_i,a_i )+\alpha_i [r_i(s_i,a_i)+
\sum_{j=1,j\neq i}^n f(i,j)r_j(s_i,a_i )+
\gamma max\left(Q_i(s_i',a_i')-Q_i(s_i,a_i)\right)]
\end{eqnarray}

where $\alpha$ and $\gamma$ are learning rate and discount factor, respectively. $s_i$ and $a_i$ are current state of traffic environment and current action, respectively. $s_i'$ is its next state, $n$ is the number of traffic signal control agents surrounding $i$-th agent, $Q_i(s_i,a_i)$ is the Q-value function for $i$-th agent when selects action $a_i$ in state $s_i$. $r_i(s_i,a_i)$ is reward function of $i$-th agent and $r_j(s_i,a_i)$ is reward function of $j$-th agent neighboring $i$-th agent. $f(i,j)\in[0,1]$ is a weighted function which shows the effect of $r_j(s_i,a_i)$ on $i$-th agent. Mathematical functions are suggested in ~\cite{XL2009IFCSTA} for $r(s,a)$ and $f(i,j)$.
Assumption of discrete action-state space and determination of reward and weighting functions are drawbacks of that work.

\section{Problem Statements}
Consider a traffic network in which the lights of each intersection is controlled by an autonomous agents without any centeralized management. Some sensors which are installed below the surface of surrounding streets or traffic cameras of each intersection provide information about traffic situation for the corresponding agent. An agent has to decide on duration of green light at North-South (NS) and West-East (WE) paths. Also, any agent interacts with neighbour agents. Anyway, the agent is expected to schedule traffic lights optimally, in the sense of average delay, based on the received information from its sensors and received information from neighbor agents. 
%This scenario with 5 intersection is depicted in Fig.\ref{Traffic}. 

The agents may have little knowledge about others' decision due to distribution of information. Even if an agent has previous known information about others' decision, it is not valid as other agents are also learning. Thus, the environment is dynamic and the behavior of other agents may change during time. Lack of prediction of other agents causes uncertainty in problem solving procedure. This paper looks for a decision-making algorithm for lights control agents which considers neighbour agents information in addition to its own information.

\section{Proposed algorithm}
\begin{figure}[!h]
	\includegraphics[width=8.3cm]{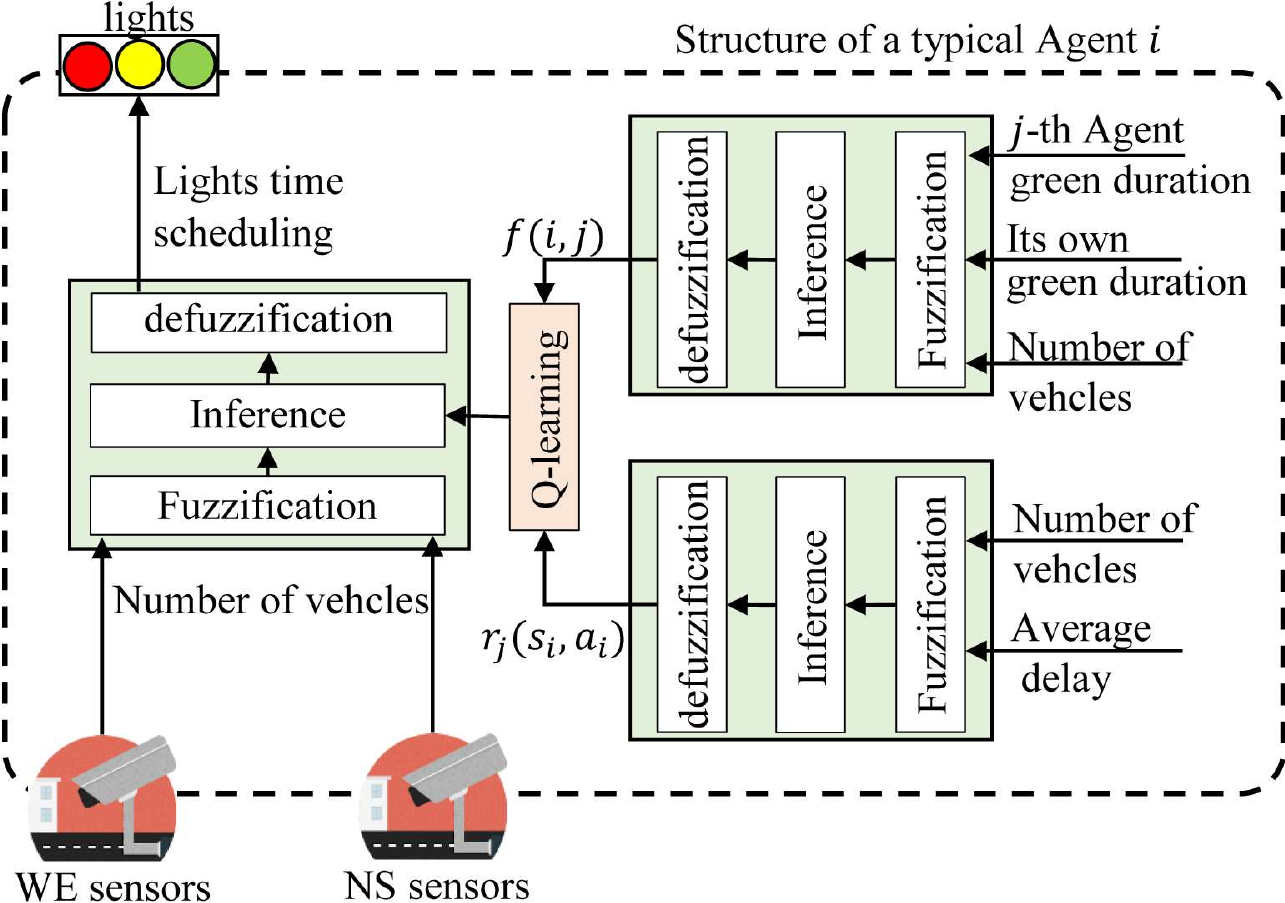}
	\caption{The proposed structure for a typical agent}
	\label{Agent}
\end{figure}

We consider a constant duration $T$ for green plus red phases. So, if the agent determines the green phase duration $t_g$, then the red phase duration is $t_r=T-t_g$. Any typical agent $i$ receives number of vehicles on the NS and WE streets from its own sensors and the green phase duration of neighbour agent $j$ in order to schedule its own green phase duration. This paper proposes an autonomous agent with structure in Fig.\ref{Agent} to control each intersection.

The number of vehicles in WE and NS streets which are measured by sensors are fuzzified. Then, a fuzzy inference engine with rules as Eq.\ref{FuzzyQLearningRule} are employed to fire the corresponding output membership functions. Finally, defuzzification results to duration of green phase in NS path ($t_g^{NS}$). Thus, the duration of green phase in other path, WE, is $t_g^{WE}=T-t_g^{NS}$.
We propose that, Q-value function which is updated by Eq.\ref{GQLearning} be the value of each action in Eq.\ref{FuzzyQLearningRule} which is denoted by $q[i,j]$. This update equation takes the neighbour agents' decision into account.

The $i$-th agent takes decision of neighbor agent $j$ into account by reward $r_j(s_i,a_i)$ and a weighting function $f(i,j)$. 
The reward is calculated based on average delay obtained from the decision made by the agent and current traffic situation in a fuzzy manner. A fuzzy inference engine obtains these two inputs after fuzzification and gives the reward after defuzzification; see Fig.\ref{Agent}. 
weighting function $f(i,j)\in[0,1]$ shows the effect of $r_j(s_i,a_i)$ on the decision of $i$-th agent. This weight is also calculated by a fuzzy inference engine. This engine takes its own $t_g$, the neighbour agents' $t_g$, and number of waited vehicles and gives $f(i,j)$.
Suitable choice for reward and weighting function plays a significant role in agent learning.
The agent with structure in Fig.\ref{Agent} runs the following algorithm:
\begin{enumerate}
  \item
        Initial value of $Q_i$-value for i-th traffic signal control agent is in the form of $\forall(s_i,a_i ):Q_i(s_i,a_i)=0$.
  \item
        Observing $s_i$ by WE and NS sensors which is the current state of $i$-th intersection.
  \item
        Selecting a proper estimation for desired state by fuzzy inference system.
  \item
        Calculating the reward related to $i$-th and $j$-th traffic signal control agent and the weighting function for neighboring agents separately.
  \item
        Observing new state $s_i'$.
  \item
        Updating $Q_i$-value according to equation \ref{GQLearning}.
  \item
        Returning to step 2 till the variation of Q-value becomes less than $\epsilon$.
\end{enumerate}
	
\section{Simulation results}
Consider a traffic network with a center and four neighbor intersection. The delay in each intersection depends on physical characteristics of the intersection, traffic light scheduling and number of cars in input streets.
We utilized traffic model which is given by the American Highway Capacity Manual (HCM) ~\cite[Eq.20]{AB1999ITE}:
\begin{eqnarray}\label{EqTrafficModel}
  d&=&0.38\frac{C(1-\lambda)^2}{1-\lambda x}+173 x^2\left[(x-1)+\sqrt{(x-1)^2+\frac{16 x}{C}}\right]
\end{eqnarray}
where $d$, $C$, $\lambda$, and $x$ are average delay (sec), cycle time (sec), green ratio, and degree of saturation, respectively. $\lambda=\frac{g}{c}$ and $x=\frac{v}{c}$, where $c$, $g$, and $v$ are capacity (vehicle per hour), green time (sec), and input volume, respectively. We use this model to calculate average delay based on the green phase duration and number of vehicles. For more details of this equation we refer to ~\cite{AB1999ITE}. 

Assume that $C=T=100sec$ and $c=3500 veh/h$. $v$ is volume of vehicles entering each street which varies between $0$ to $3500veh/h$. $g$ is duration of the green phase which each agent selects considering fuzzy Q-learning and interaction with adjacent agents. The traffic network simulation algorithm is as follow:
\begin{enumerate}
	\item
	The volume of vehicles entering each intersection ($v$) are randomly generated by a discrete uniform distribution on the interval $[0,3500]$. 
	\item
	Average delay is calculated by Eq.\ref{EqTrafficModel}.
	\item
	Each agent decides on the time of green phase $g$.
	\item
	Go to step 1 until end of simulation time.
\end{enumerate}

\begin{figure}[!h]
	\includegraphics[width=6cm]{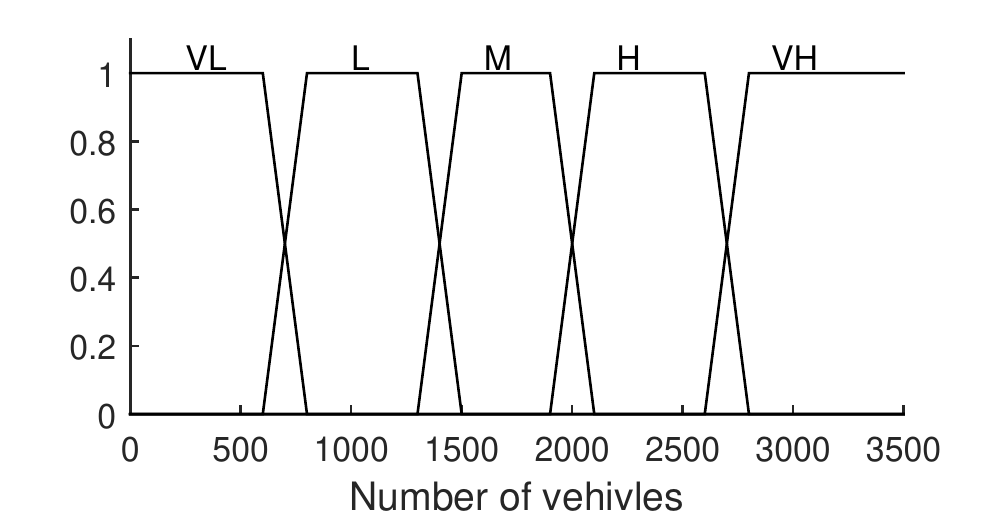}
	\caption{Membership function of number of vehicles enter the street for reward FIS}\label{MFIN1REW}
\end{figure}

\begin{figure}[!h]
	\includegraphics[width=6cm]{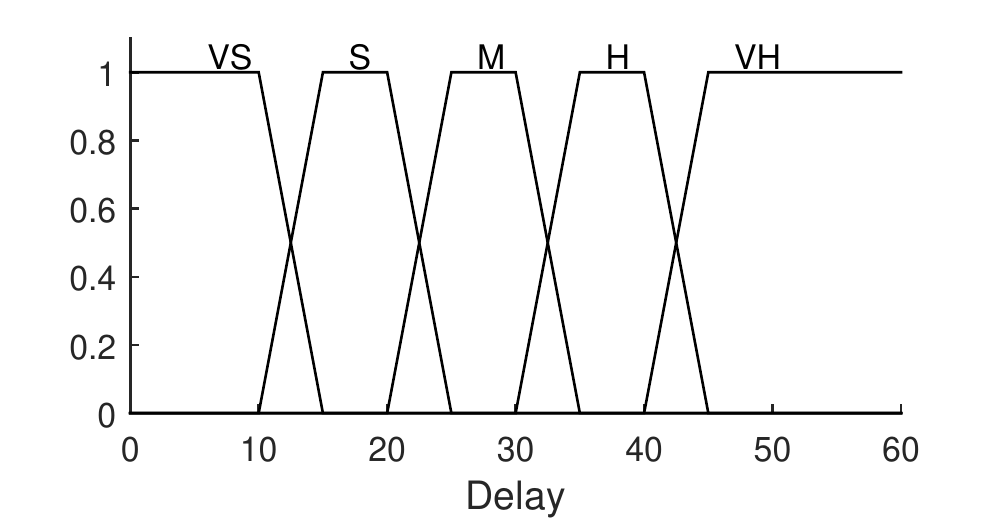}
	\caption{Membership function of average delay for reward FIS}\label{MFIN2REW}
\end{figure}

\begin{figure}[!h]
	\includegraphics[width=6cm]{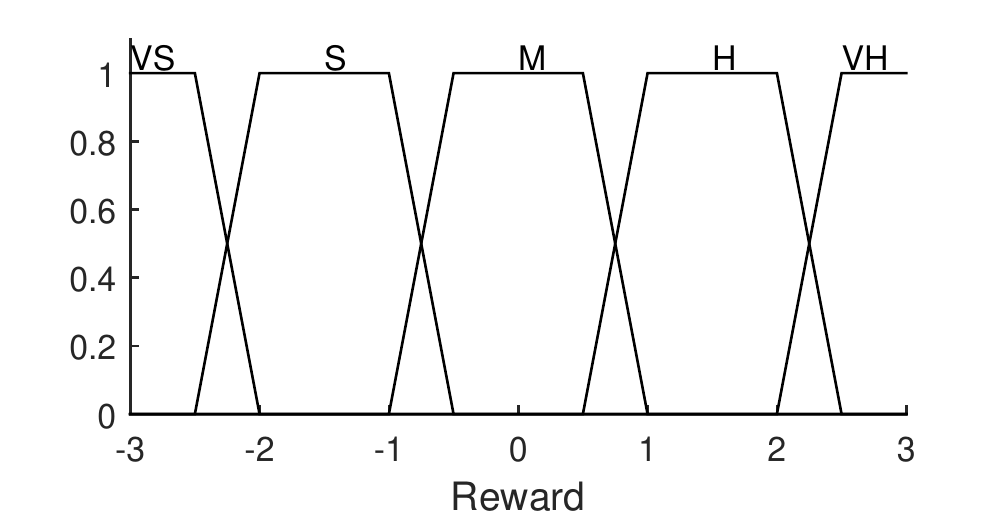}
	\caption{Membership function of output for reward FIS}
	\label{MFOUTREW}
\end{figure}

Assume structure of the agents as in Fig.\ref{Agent} with the Mamdani FIS with input membership function as in Fig.\ref{MFIN1REW} for number of input vehicles and Fig.\ref{MFIN2REW} for average delay to calculate the reward functions $r_j(s_i,a_i)$. Centroid defuzzification by the output membership function as in Fig.\ref{MFOUTREW} is considered to estimate a reward value in interval $[-3,3]$.


\begin{figure}[!h]
	\includegraphics[width=6cm]{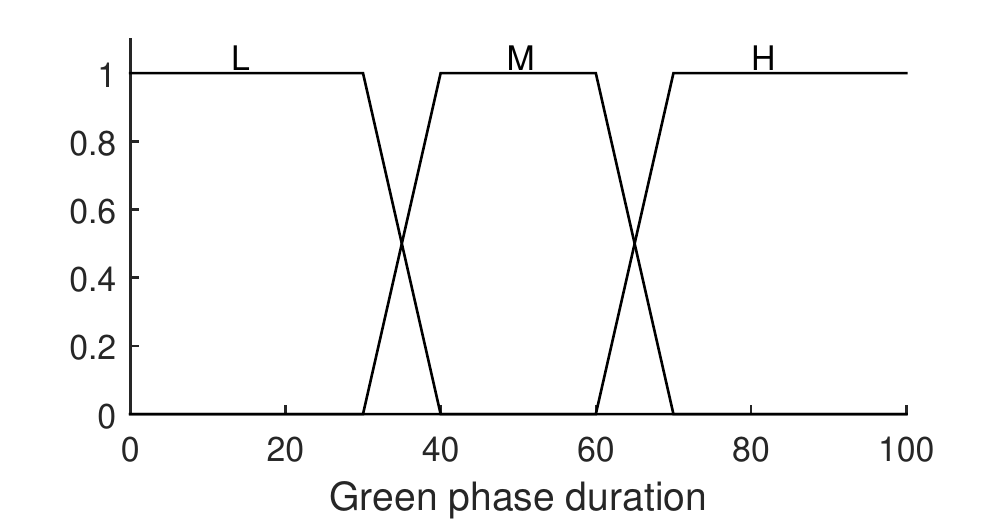}
	\caption{Membership function of green phase duration for weighting function FIS}
	\label{MFIN2WF}
\end{figure}

\begin{figure}[!h]
	\includegraphics[width=6cm]{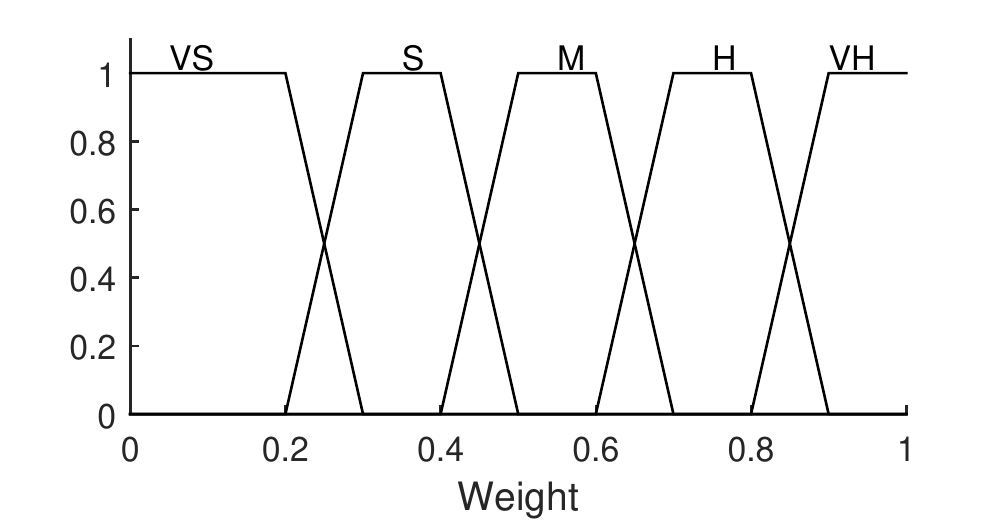}
	\caption{Membership function of output for weighting function FIS}
	\label{MFOUTWF}
\end{figure}

The weighting function FIS has number of vehicles, its own green phase duration and the neighbour agents' green phase duration as inputs. Fig.\ref{MFIN1REW} shows the membership function for number of vehicles and Fig.\ref{MFIN2WF} depicts the membership function for its own and neighbour green phase duration. Centroid defuzzification is applied to calculate weights on output membership function as in Fig.\ref{MFOUTWF} which should be a value between $0$ and $1$.


\begin{figure}[!h]
	\includegraphics[width=6cm]{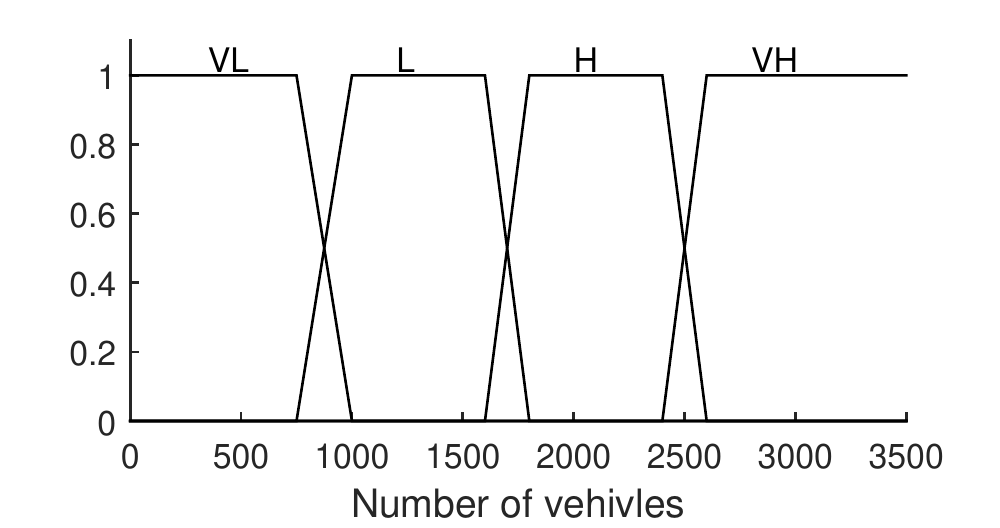}
	\caption{Membership function of number of vehicles for fuzzy Q-learning}
	\label{MFINFQL}
\end{figure}

\begin{figure}[!h]
	\includegraphics[width=6cm]{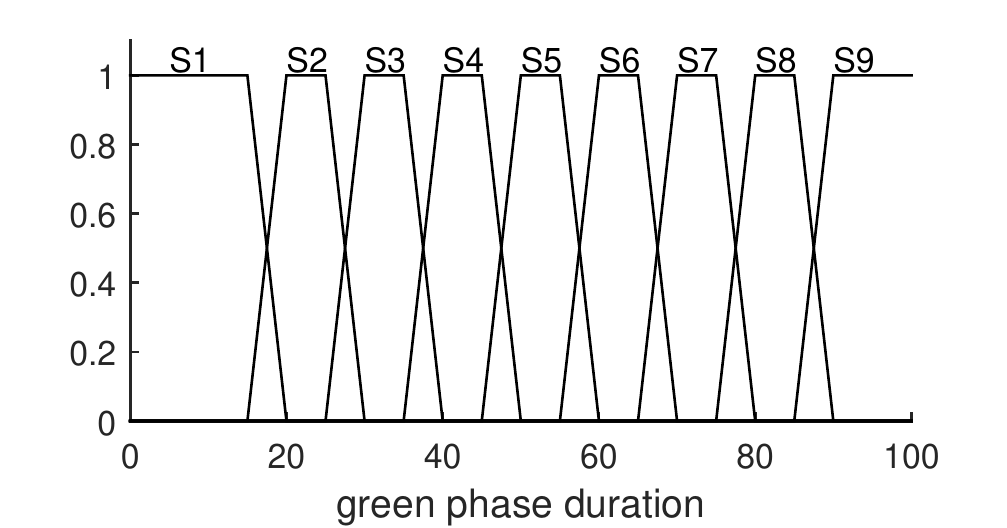}
	\caption{Membership function of green phase duration for fuzzy Q-learning}
	\label{MFOUTFQL}
\end{figure}

Finally, the agent uses fuzzy Q-learning (Eq.\ref{FuzzyQLearningRule}) with Q-value update rule (Eq.\ref{GQLearning}) where learning and discount factor are selected to be 0.5 and 0.7, respectively. The membership function for each measured number of vehicles is shown in Fig.\ref{MFINFQL}. The output estimates green phase duration with membership functions as in Fig.\ref{MFOUTFQL}.


\begin{figure}[!h]
	\includegraphics[width=8.3cm]{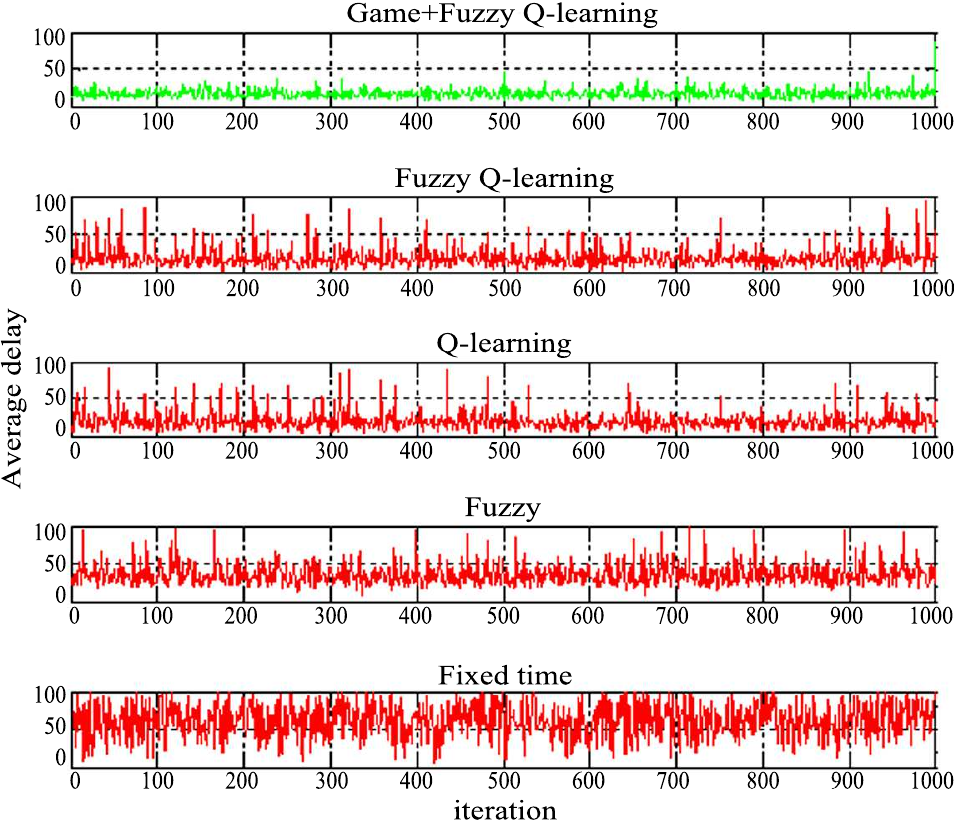}
	\caption{Delay of the proposed method, fixed time, fuzzy Q-learning, Q-learning and fuzzy in each time step}\label{delay}
\end{figure}

\begin{figure}[!h]
	\includegraphics[width=7.8cm]{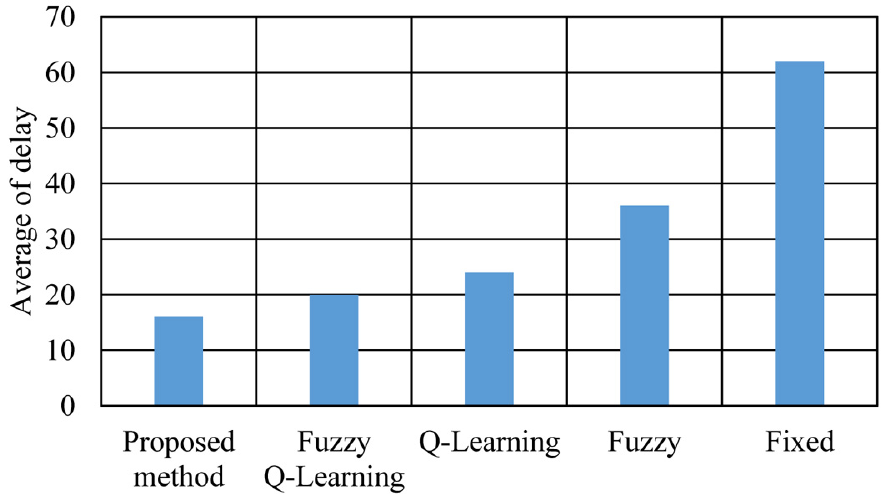}
	\caption{Average of delay for the proposed method, fixed time, fuzzy, Q-learning, fuzzy Q-learning}\label{Compare}
\end{figure}

The proposed method is compared with Fuzzy Q-learning (using Eq.\ref{FuzzyQLearningRule} where $q[i,j]$ is the Q-value which updates with Eq.\ref{QLearning}), Q-learning (using Q-learning method with Q-value which updates with Eq.\ref{QLearning}), fuzzy(using traditional fuzzy inference method) and fixed time ($t_g=60sec$) in the sense of total average delay. Average delay in each time interval is depicted in Fig.\ref{delay} and the total average delay is illustrated in Fig.\ref{Compare}. The results illustrate that total average delay decrease from more than $50sec$ for fixed time scheduling to approximately $15sec$ for the proposed method.


\section{Conclusion}
In this study an intelligent control method of a controlling traffic network was performed to decrease average delay time. Each traffic light is considered as a learning agent. This paper proposed a structure for the agents. Each agent learn to decide on the duration of green phase through a fuzzy Q-learning algorithm which is modified by Game theory. Each agent receives a reward from neighbour agents. The reward received from the neighbour and weighted functions of neighboring agents are factors learning algorithm. These parameters are fuzzified through a FIS. Also, the number of vehicles in each street is measured and fuzzified to be used in decision making process. The simulation results were compared with fixed time method and other intelligent methods. The results revealed that our proposed method achieves considerable reduction of average delay in intersections.

\bibliographystyle{spbasic}     % basic style, author-year citations
%\bibliographystyle{splncs}
%\bibliographystyle{unsrtnat}
%\bibliography{Citations}

\end{document}